\documentclass[aps,superscriptaddress,twoside,onecolumn,floatfix,a4paper,longbibliography]{revtex4-2}

\usepackage{graphicx}
\usepackage{amsmath}
\usepackage{dcolumn}   
\usepackage{bm}        
\usepackage{amssymb}   
\usepackage{graphicx,epsfig}
\usepackage{color}
\usepackage[usenames,dvipsnames]{xcolor}
\usepackage[ocgcolorlinks,colorlinks=true,linkcolor=blue,citecolor=red]{hyperref}
\usepackage{amsmath,amssymb, amsthm, amsfonts}
\usepackage{xspace}
\usepackage{xcolor}
\usepackage{verbatim}
\usepackage{mathtools}
\usepackage{lipsum}
\usepackage{natbib}
\definecolor{goodgreen}{rgb}{0.1,0.5,0}

\usepackage[utf8]{inputenc}

\begin{document}
\title{Bipolar Thermoelectric Josephson Engine}

\author{Gaia Germanese}
\affiliation{NEST, Istituto Nanoscienze-CNR and Scuola Normale Superiore, I-56127 Pisa, Italy}
\affiliation{Dipartimento di Fisica dell’Università di Pisa, Largo Pontecorvo 3, I-56127 Pisa, Italy}

\author{Federico Paolucci}
\email{federico.paolucci@sns.it}
\affiliation{NEST, Istituto Nanoscienze-CNR and Scuola Normale Superiore, I-56127 Pisa, Italy}

\author{Giampiero Marchegiani}
\affiliation{Quantum Research Centre, Technology Innovation Institute, Abu Dhabi, UAE}

\author{Alessandro Braggio}
\affiliation{NEST, Istituto Nanoscienze-CNR and Scuola Normale Superiore, I-56127 Pisa, Italy}

\author{Francesco Giazotto}
\email{francesco.giazotto@sns.it}
\affiliation{NEST, Istituto Nanoscienze-CNR and Scuola Normale Superiore, I-56127 Pisa, Italy}

\begin{abstract}
Thermoelectric effects in metals are typically small due to the nearly-perfect particle-hole (PH) symmetry around their Fermi surface.
Furthermore, thermo-phase effects and linear thermoelectricity in superconducting systems were identified only when PH symmetry is explicitly broken, since thermoelectric effects were considered impossible in pristine superconductors. 
Here, we experimentally demonstrate that superconducting tunnel junctions develop a very large bipolar thermoelectricity in the presence of a sizable thermal gradient thanks to \textit{spontaneous} PH symmetry breaking. 
Our junctions show Seebeck coefficients up to $\pm$300$\;\mu$V/K, that is comparable with quantum dots and roughly 10$^5$ times larger than the value expected for normal metals at subKelvin temperatures. Moreover, by integrating our junctions into a Josephson interferometer, we realize a bipolar thermoelectric Josephson engine (BTJE) generating phase-tunable electric powers up to $\sim 140$ nW/mm$^2$.
Notably, our device implements also the prototype for a persistent thermoelectric memory cell, written or erased by current injection.
We expect that our findings will lead to applications in superconducting quantum technologies.
\end{abstract}

\maketitle

Thermoelectricity is the capability of materials to directly convert temperature gradients into an electrical power \cite{Ashcroft1976, Abrikosov1988}. Specifically, a thermoelectric element can supply a short-circuit current (\textit{Peltier regime}) or generate an open-circuit voltage (\textit{Seebeck regime}) whose sign is determined  by the sign of the dominant carriers and temperature gradient.
All systems characterized by strong PH symmetry, such as normal metals and superconductors, show poor thermoelectric effects \cite{Mott1958,vanHarlingen1984}. Moreover, thermoelectricity in superconductors is also screened by the dissipationless motion of Cooper pairs \cite{Meissner1927}, thereby only thermo-phase effects can be eventually observed \cite{Ginzburg1944,Shelly2016,Giazotto2015,Kleeorin2016,Guttman1997}. Yet, a pure local thermoelectric effect can only be generated in superconducting tunnel junctions with suppressed Josephson coupling by explicitly breaking the PH symmetry \cite{Smith1980,Ozaeta2014,Machon2013,Bergeret2018,Linder,Kolenda2016}, whereas non-local thermoelectricity can be detected in  superconducting hybrid structures \cite{Virtanen2004,Blasi2020,Tan2021,eom,chandrasekar,Hofstetter}. 
These intrinsic limitations hindered so far the implementation of superconducting thermoelectric devices in quantum technologies, such as radiation detectors, switches, memories, and engines. 
Indeed, despite great theoretical efforts \cite{Campisi2015,Benenti2017,Bera2021}, the experimental realization of efficient solid-state heat engines is still limited to InAs/InP quantum-dots \cite{Josefsson2018}, molecular systems \cite{Dubi2011} and silicon tunnel transistors \cite{Ono2020}.

Here, we report the experimental observation of an astounding bipolar thermoelectric effect in tunnel junctions between two different superconductors subject to nonlinear thermal gradients \cite{Marchegiani2020}, i.e.,  when the temperature gradient imposed across the junction is comparable  to its average value (nonlinear response regime).
Our junctions show a maximum thermovoltage of $\pm$150 $\mu$V at $\sim$650 mK, directly proportional to the superconducting gap. 
The corresponding Seebeck coefficient $\mathcal{S}\simeq\pm$300$\;\mu$V/K is roughly 10$^5$ times larger than the one expected for normal metals forming the structures \cite{Mott1958,vanHarlingen1984} at the same temperature. Notably, these values of $\mathcal{S}$ are two orders of magnitude larger than conventional bulk thermoelectric materials \cite{Roddaro2013,Soleimani2020} and compare with quantum dots \cite{Mani2009,Prete2019} at cryogenic temperatures.  

The unique bipolarity of the reported effect stems from the equal possibility for \emph{both} directions of the thermocurrent or polarities of the thermovoltage at a \emph{given} temperature gradient and electron configuration in the superconducting electrodes.
The phenomenology is determined by nonequilibrium \emph{spontaneous} breaking of  PH symmetry and can be phase-controlled in a Josephson interferometer \cite{Marchegiani2020_3}.
We exploit these features to realize a bipolar thermoelectric Josephson engine (BTJE) \cite{Marchegiani2020_2}.
When connected to a generic load, the BTJE generates a phase-tunable electric power up to $\sim 140$ nW/mm$^2$ at subKelvin temperatures.
Finally, we expect that the BTJE may find immediate application in superconducting quantum technology \cite{Patent}.

\subsubsection*{\textbf{Design of the Bipolar Thermoelectric Josephson Engine}}
\begin{figure*}[t!]
	\centering 
	\includegraphics[width=0.75\linewidth]{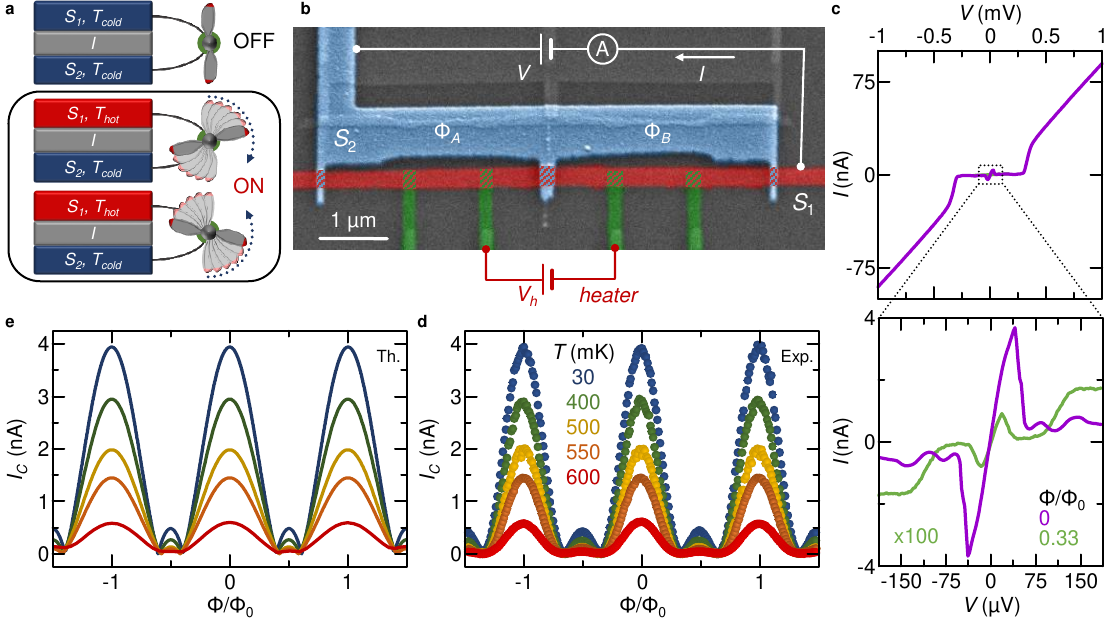}
	\caption{\textbf{Bipolar thermoelectric Josephson Engine.}
	\textbf{a} Scheme of the bipolar thermoelectric Josephson engine (BTJE): two different superconductors S$_1$ and S$_2$ (with zero-temperature energy gaps $\Delta_{0,1} > \Delta_{0,2}$) are tunnel-coupled through a thin insulating layer (grey, I). The S$_1$IS$_2$ junction is predicted to generate power  when S$_1$ is kept at a higher temperature than S$_2$ (i.e., $T_1=T_{hot} > T_2= T_{cold}$). Remarkably, this structure can produce both \textit{positive} and \textit{negative} thermovoltage for the same thermal gradient imposed across the junction.
	\textbf{b} Pseudo-color scanning electron micrograph of a typical BTJE. An aluminum island (S$_1$, red) is tunnel coupled through three AlO$_x$ barriers to a Cu/Al bilayer (S$_2$, blue) thus realizing a double-loop SQUID. Additional tunnel-coupled Al electrodes (green) serve as Joule heaters for S$_1$. Replicas resulting from the shadow-mask fabrication procedures are also visible (see Methods for details). The charge transport properties of the double-loop SQUID were investigated in a two-wire configuration by recording the tunneling current ($I$) while a voltage ($V$) was applied. $\Phi_A$ and $\Phi_B$ represent the magnetic fluxes linked to each loop, so that the total flux through the interferometer is $\Phi=\Phi_A + \Phi_B$. 
	\textbf{c} Top panel: current ($I$) versus voltage ($V$) characteristics measured at $30$ mK for $\Phi=0$ (violet) and $\Phi=0.33\Phi_0$ (green). Bottom panel: blow-up of the $IV$ characteristics around $V=0$.
	\textbf{d} Experimental Josephson critical current ($I_C$) of the double-loop interferometer vs $\Phi$ measured for different values of bath temperature. \textbf{e} Theoretical modulations of $I_C$ vs $\Phi$ calculated with the model presented in the SI.}
	\label{Fig1}
\end{figure*}

The core of the BTJE is a S$_1$IS$_2$ tunnel junction between two different Bardeen-Cooper-Schrieffer superconductors with suppressed Josephson coupling (where S$_1$ and S$_2$ have  zero-temperature energy gaps $\Delta_{0,1}>\Delta_{0,2}$, and I stands for an insulator).
At thermal equilibrium (i.e., for identical temperatures $T_1=T_2=T_{cold}$), the junction is dissipative, and cannot generate power \cite{Benenti2017} (see top panel of Fig. \ref{Fig1}a). 
By contrast, in the presence of a suitable thermal bias (which approximately satisfies $T_1\gtrsim T_2\Delta_{0,1}/\Delta_{0,2}$), the junction is expected to yield thermoelectricity with optimal performance for $r=\Delta_{0,2}/\Delta_{0,1}=0.2...0.5$. \cite{Marchegiani2020} (see Methods for details). 
As we shall show, this structure may spin an electrical motor in both directions for a \emph{given} thermal gradient and electron configuration in the superconductors depending on the polarization history of the junction, as sketched in the bottom panel of Fig.~\ref{Fig1}a. By contrast, the sign of the thermocurrent generated by quantum dot-based engines is determined by their electronic configuration, and can be controlled through electrostatic gating \cite{Josefsson2018}.


Our implementation of the BTJE is shown by the false-color scanning electron micrograph in Fig.~\ref{Fig1}b. It consists of a double-loop superconducting quantum interference device (SQUID) \cite{Kemppinen2008,Fornieri2016}, where S$_1$ (red, Al) is coupled to S$_2$ (blue, Al/Cu bilayer) through three insulating AlO$_x$ tunnel junctions. 
In this configuration, the three S$_1$IS$_2$ junctions constitute a Josephson interferometer, which is used to  phase-control the thermoelectric generation \cite{Marchegiani2020_3} via fine tuning of the dissipationless supercurrent.
In particular, a double-loop SQUID guarantees a more effective suppression of the Josephson coupling with respect to a conventional single-loop two-junction interferometer \cite{Kemppinen2008,Fornieri2016}, thereby allowing an improved control of the thermoelectric effect. 
Moreover, S$_1$ is also equipped with several superconducting tunnel junctions (green, Al) operated as Joule heaters to establish the necessary temperature gradient across the structure. The structure fabrication details are provided in the Methods section.

Investigation of charge transport in the BTJE is performed first in the absence of a temperature gradient (i.e., for $T_1=T_2=T$).
Figure \ref{Fig1}c shows the 2-wire SQUID current ($I$) versus voltage ($V$) characteristics measured at $30$ mK for two representative values of magnetic flux ($\Phi$). The SQUID switches to the normal-state characteristic at $V=\pm (\Delta_{0,1}+\Delta_{0,2})/e\simeq \pm 300$ $\mu$V (with $e$ the electron charge) displaying a total junction resistance $R_T\sim9$ k$\Omega$ (top panel).
Moreover, the Josephson critical current, manifesting itself as a peak around zero bias, is significantly modulated by $\Phi$ (bottom panel). 
The interference patterns of the critical current ($I_C$) recorded at different values of bath temperature are shown in Fig.~\ref{Fig1}d. 
In agreement with our model (see Fig. \ref{Fig1}e, and SI for the model details), the double-loop geometry allows an effective and fine phase-tuning of the SQUID transport properties, and a maximum suppression of the supercurrent up to $\sim 1.75$\textperthousand\;  of the zero-flux value for $\Phi=0.33\Phi_0$.


\begin{figure*}[ht!]
	\centering 
	\includegraphics[width=0.75\linewidth]{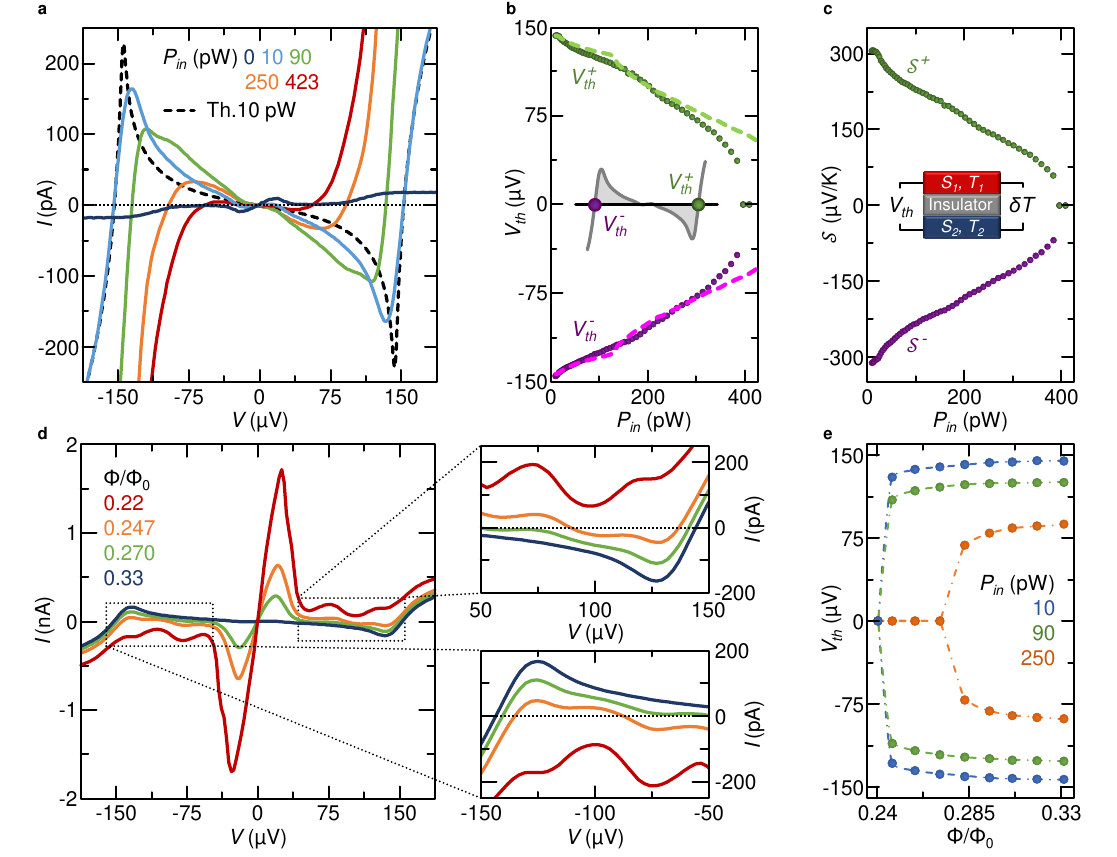}
	\caption{\textbf{Bipolar thermoelectric effect.}
	\textbf{a} Subgap current ($I$) vs voltage ($V$) characteristics of the SQUID measured at $\Phi=0.33\Phi_0$ for different values of the input power ($P_{in}$) injected in S$_1$. In the presence of a thermal bias, the interferometer shows an \textit{absolute negative}
	conductance (ANC) for both polarities of $V$. The black dashed line is the result from the theory for a single junction \cite{Marchegiani2020} calculated at $10$ pW. 
	The temperature bias across the BTJE is deduced by  fitting the experimental $IV$ curves with the theory (see SI for details).
	\textbf{b} Thermovoltage ($V_{th}$) vs $P_{in}$ recorded at $\Phi=0.33\Phi_0$ (dots). Dashed lines correspond to the theory \cite{Marchegiani2020}. Due to intrinsic electron-hole symmetry, the system provides both \textit{positive} ($V_{th}^+$, green) and \textit{negative} ($V_{th}^-$, purple) thermovoltage for any given $P_{in}$.
	\textbf{c} Seebeck coefficient ($\mathcal{S}$) vs $P_{in}$ extracted from the data shown in panel b according to the calibration presented in the SI. In the presence of thermoelectricity, the value of $\mathcal{S}$ monotonically decreases by increasing $P_{in}$.
	\textbf{d} Subgap $IV$ characteristics of the SQUID measured at $P_{in}=10$ pW for selected values of $\Phi$. Top (bottom) inset shows a magnification of the current close to $V_{th}^+$ ($V_{th}^-$).
	\textbf{e} $V_{th}$ vs $\Phi$ measured for different values of $P_{in}$. Dash-dotted lines are guides to the eye.
	All measurements were performed at a bath temperature of $30$ mK.}
	\label{Fig2}
\end{figure*}

\begin{figure*}[t!]
	\centering 
	\includegraphics[width=0.75\linewidth]{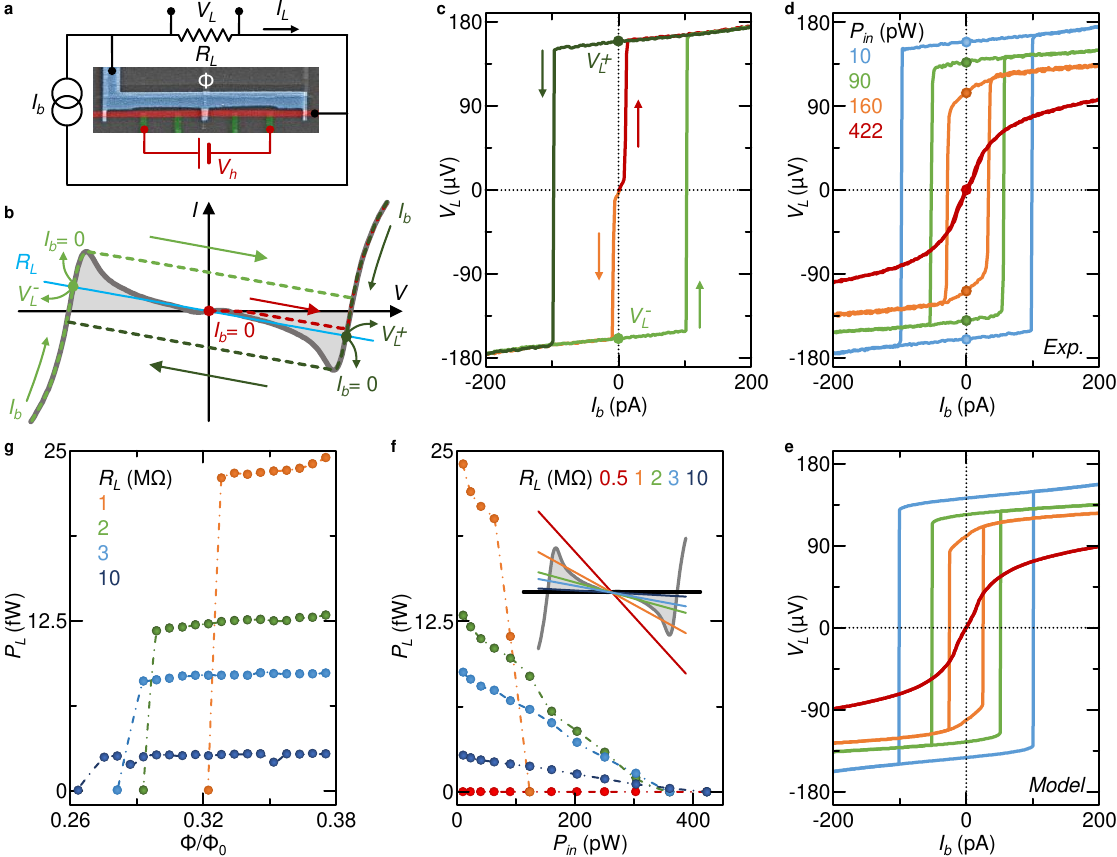}
	\caption{\textbf{Low temperature behavior of the Bipolar Thermoelectric Josephson Engine.}
	\textbf{a} Scheme of the electronic circuit used to demonstrate power production.
	The Joule heaters are powered through a floating voltage source ($V_h$).
	A dc current source ($I_b$) biases the parallel connection of the interferometer and a load resistor ($R_L$)  while recording the voltage drop $V_L$ occurring across $R_L$.
	\textbf{b} Sketch of thermoelectric current ($I$) vs voltage ($V$) characteristic where a resistive load (blue line) is superimposed. A typical hysteresis loop of the load voltage ($V_L$) is also represented (green dashed curves) together with the thermovoltages ($V_L^+$ or $V_L^-$) corresponding to zero biasing  current (i.e., $I_b=0$). Grey areas represent \textit{thermoactive} regions.
	\textbf{c} $V_L$ vs $I_b$ recorded with $R_L=2$ M$\Omega$ at $P_{in}=10$ pW and $\Phi=0.33\Phi_0$. The engine can be ignited by ramping $I_b$ from zero towards either positive (red) or negative (orange) values. Thermovoltages measured at $I_b=0$ (green dots, $V_L^+$ or $V_L^-$) prove power production in the system, i.e., the realization of the \textit{engine}. 
	The thermovoltage polarity can be changed by ramping $I_b$ to large values (green curves).  
	\textbf{d} $V_L$ vs $I_b$ characteristics measured with $R_L=2$ M$\Omega$ and $\Phi=0.33\Phi_0$ for selected values of input power.
	\textbf{e} Results from the model for the same parameters as in panel d (see SI for details).
	\textbf{f} Engine output power ($P_L$) vs input power ($P_{in}$) at $\Phi=0.33\Phi_0$ for different values of $R_L$. Inset: characteristics of different loads (colored lines) superimposed to a typical BTJE thermoactive $IV$ curve (grey).
	\textbf{g}  $P_L$ vs  $\Phi$ at $P_{in}=10$ pW for different values of $R_L$.
	Dash-dotted lines in panels f and g are guides to the eye.
	All measurements were performed at a bath temperature of $30$ mK.}
	\label{Fig3}
\end{figure*}

\subsubsection*{\textbf{Bipolar thermoelectric effect}}
To assess the predicted bipolar thermoelectric effect \cite{Marchegiani2020}, we measured the subgap $IV$ characteristics of the BTJE at a bath temperature of $30$ mK while directly injecting power ($P_{in}$) in S$_1$. Power injection raises the temperature of $S_1$ above the bath value, thereby generating a thermal gradient across the interferometer (i.e., $\delta T=T_1-T_2>0$) \cite{Giazotto2006,Fornieri2017}.
Since the Josephson coupling is detrimental for thermoelectricity \cite{Marchegiani2020_2}, we flux-biased the SQUID at $\Phi=0.33\Phi_0$ in order to minimize the supercurrent flowing through the interferometer i.e., $I_C(\Phi=0.33\Phi_0)\simeq7$ pA]. Therefore, the Josephson contribution to charge transport can be neglected in this magnetic flux condition \cite{Marchegiani2020_3}.

For nonzero input power (i.e., for $P_{in}\geq10$ pW, see SI for details about the input power generation), the subgap quasiparticles current flows against the bias voltage ($IV<0$), thus displaying the \textit{absolute negative} conductance (ANC) \cite{Spivak,Ger1}, which signals thermoelectric generation (see Fig. \ref{Fig2}a) \cite{Marchegiani2020}.
Differently from other known thermoelectric effects \cite{Ashcroft1976}, the BTJE shows \textit{bipolar} power generation, i.e., the ANC appears for both positive and negative values of $V$ at a \emph{given} temperature gradient without any external action on the electronic configuration of the superconductors. This unique anti-symmetric thermoelectric $IV$ characteristics [$I(-V)=-I(V)$] stem from PH symmetry of the two superconducting leads. 
In general, the experimental $IV$ traces are in agreement with  the  theoretical behavior obtained by computing the quasiparticle current (black dashed line) in the presence of a nonlinear temperature gradient \cite{Marchegiani2020} (see Methods and SI for details).
When the input power is too large (i.e., for $P_{in}\gtrsim400$ pW), the thermoelectricity vanishes, and the $IV$ characteristics show the conventional dissipative behavior (see the red curve in Fig. \ref{Fig2}a).
This power dependence is highlighted by the thermovoltage ($V_{th}$), i.e., the nonzero potential drop occurring across the BTJE when $I(V_{th})=0$. 
For linear thermoelectricity this quantity corresponds to the Seebeck voltage \cite{Benenti2017}.
In our measurements, the thermovoltage obtains values as large as $\sim\pm 150\;\mu$V for the lowest heating power of $10$ pW, in good agreement with  theory (see dashed line in Fig. \ref{Fig2}b). 
In contrast to  linear thermoelectricity, by increasing  $P_{in}$ (and therefore at larger temperature gradients) yields a monotonic reduction of
the absolute value of both polarities of $V_{th}$, until reaching full suppression around $400$ pW. 

It is common to evaluate the performance of a thermoelectric element
through the Seebeck coefficient, defined as $\mathcal{S}=V_{th}/\delta T$.
The resulting bipolar Seebeck coefficient vs $P_{in}$ is shown in Fig. \ref{Fig2}c (see SI  for details). 
For the BTJE, $\mathcal{S}$ can be as high as $\pm300\;\mu$V/K at $P_{in}=10$ pW (corresponding to a temperature of $\simeq650$ mK in S$_1$), which is near to the maximum theoretically achievable with our materials \cite{Marchegiani2020,Marchegiani2020_2}.
The above value is almost $10^5$ times larger than the Seebeck coefficient of aluminum [$\mathcal{S}_{Al}(650\;\text{mK})\simeq-3.8$ nV/K] \cite{Mott1958,vanHarlingen1984}, which is the main constituent of our structure, as computed through the Mott-Jones equation (see Methods for details).
Our system compares well with quantum dots, which are currently the best thermoelectric materials at cryogenic temperatures providing Seebeck coefficients of the order of several hundred $\mu$V/K for temperatures up to $\sim 30$ K \cite{Mani2009,Prete2019}. Instead, more conventional thermoelectric materials for room temperature applications are characterized by values of $\mathcal{S}$ linearly decreasing with temperature, and reaching only about 1 $\mu$V/K at 1 K \cite{Roddaro2013,Soleimani2020}.

To demonstrate the interplay between our thermoelectric effect and the Josephson coupling \cite{Marchegiani2020_3}, we measured the $IV$ characteristics at a given input power ($P_{in}=10$ pW) for different values of the magnetic flux piercing the interferometer.
As expected, the thermoelectric effect is strongly suppressed for both voltage polarities in the presence of a sizeable Josephson current  (see Fig. \ref{Fig2}d). 
Specifically, at low $P_{in}$,  thermoelectricity vanishes for $\Phi\simeq 0.24\Phi_0$,  corresponding to $I_C\simeq850$ pA (see Fig. \ref{Fig2}e). 
This behavior proves full $\Phi$-control of $V_{th}$  until its complete suppression \cite{Marchegiani2020_3} and is unique to the BTJE. 
It is noteworthy that Josephson coupling affects more easily thermoelectricity when it is weaker, as is happens at larger injected power (see orange curve in Fig. \ref{Fig2}e). 
As discussed above, $P_L$ is reduced by raising $P_{in}$, due to the peculiar nonlinearity of the thermoelectric effect. As a result, even small values of the Josephson current shunting the interferometer can suppress the thermoelectric generation at a large temperature bias.

\subsubsection*{\textbf{Operation of the Bipolar Thermoelectric Josephson Engine}}
Let us now analyze the behavior of the interferometer when performing as an engine. 
The electronic circuit used to operate the SQUID as a thermoelectric engine \cite{Marchegiani2020_2} is schematically shown in Fig. \ref{Fig3}a, where the device is connected in parallel to a load resistor ($R_L$), and biased by a dc current ($I_b$).
In the presence of a thermal gradient ($P_{in}>0$), the combined system BTJE $\parallel R_L$ exhibits  up to three metastable states, as shown by the colored dots in Fig. \ref{Fig3}b. These solutions are obtained by the intersection between the $IV$ characteristic of the SQUID (grey curve) and one of the resistive load (cyan line). Initially, the system is in the PH symmetric point, and the most favourable solution is $V=0$ (red dot). Therefore, the engine is off ($I=0$). The BTJE is ignited through a bias current ($I_b$), and its polarity can be selected by the sign of $I_b$. 
Indeed, positive (negative) values of $I_b$ rigidly shift up (down) the load curve upon the interferometer $IV$ characteristic thus selecting the positive (negative) thermoactive branch of the BTJE. After ignition, the engine yields a voltage drop $V_L^+$ (or $V_L^-$) across $R_L$ at $I_b=0$. By decreasing (increasing) the bias current, the load curve can intercept the BTJE characteristic only in the negative (positive) thermoactive branch thereby inverting the operation polarity of the engine without requiring any modification of the BTJE electronic configuration. Details of the circuit operation are provided in the SI. 

The bipolar power generation of the BTJE leads to the experimental $V_L$ vs $I_b$ traces shown in Fig. \ref{Fig3}c for $R_L=2$ M$\Omega$, $P_{in}=10$ pW and $\Phi=0.33\Phi_0$. Indeed, the engine is ignited by a positive (negative) current bias, represented with a red (orange) curve and, then, it is able to generate  positive (negative)  voltages across $R_L$ (i.e., $V_L^+$ and $V_L^-$) even when the bias is switched off $I_b=0$. 
Remarkably, by using the Bogoliubov criterium for phase transitions \cite{SSBReview,SSBBook}, the hysteretic behaviour of $V_L(I_b)$ implicitly proves that the bipolar thermoelectric effect stems from \emph{spontaneous} breaking of PH symmetry in the system.

Since thermoelectricity strongly depends on the amplitude of the thermal gradient, $P_{in}$ has a marked impact on the $V_L(I_b)$ characteristics (see Fig. \ref{Fig3}d). 
In particular, the absolute values of both the voltage developed across $R_L$ and the bias current corresponding to polarity inversion of $V_L$ are reduced by rising $P_{in}$. Yet, at large $P_{in}$ values, the hysteretic behavior disappears, and the BTJE turns off, since the thermoelectric effect vanishes in the presence of sizable temperature gradients (see Fig. \ref{Fig2}a).  
The above behavior is in full agreement with the  solution of the circuital model describing our experiment (see Fig. \ref{Fig3}e, and SI for details). We also emphasize that thanks to its  hysteretic character  this circuit directly implements a volatile thermoelectric \textit{memory cell} with the capability to be written/erased by current pulses \cite{Patent}. 


\begin{figure}[t!]
	\centering 
    \includegraphics[width=0.5\columnwidth]{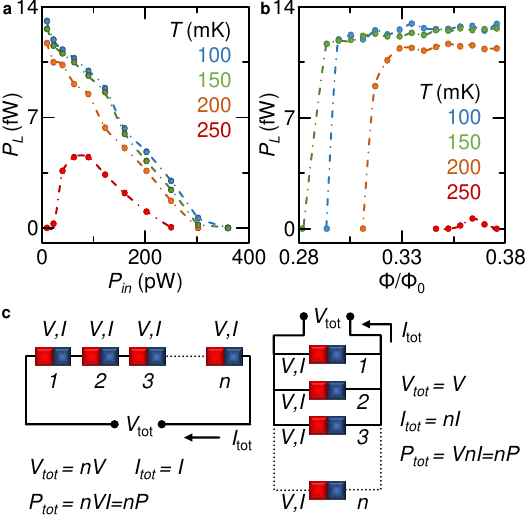}
	\caption{\textbf{Temperature dependence of the Bipolar Thermoelectric Josephson Engine.}
	\textbf{a} Output power $P_L$  at $\Phi=0.33\Phi_0$ and $R_L=2$ M$\Omega$ vs input power $P_{in}$ for different values of bath temperature $T$.
	\textbf{b} $P_L$  at $P_{in}=10$ pW and $R_L=2$ M$\Omega$     vs $\Phi$ for selected values of $T$. Dash-dotted lines in panels a and bare guides to the eye.
	\textbf{c} Schematic representation of a series (left) and parallel (right) connection of \textit{n} S$_1$IS$_2$ thermoelectric elements. The series connection yields a total voltage $V_{tot}=nV$ (with $V$ the voltage drop occurring across each element) and a total current $I_{tot}=I$ (with $I$ the current flowing through each element), whereas the parallel connection provides $V_{tot}=V$ and $I_{tot}=nI$. Therefore, both configurations generate an output power $P_{tot}=I_{tot}V_{tot}=nIV=nP$ ($P$ denotes the power produced by a single element).}
	\label{Fig4}
\end{figure}

Figure \ref{Fig3}f shows the output power, defined as $P_L=(V_L^{\pm})^2/R_L$, produced by the BTJE on different load resistors vs $P_{in}$ at $30$ mK, and $\Phi=0.33\Phi_0$. In full agreement with the thermovoltage behavior, $P_L$ decreases by rising the temperature gradient (with input power). 
The measurements show non-zero $P_L$ for the lowest experimental value of $P_{in}$, because our realization of the BTJE prevents to provide smaller heating powers and consequently to establish smaller temperature gradients (see SI for details). On the one hand, the engine produces more power for lower values of $R_L$, since $V_L$ is almost independent of $R_L$.  
In particular, the BTJE  delivers powers as large as $P_L\sim24$ fW for $R_L=1$ M$\Omega$, which corresponds to a surface power density of about 140 nW/mm$^2$. 
On the other hand, the thermal gradient (input power) operation window of the BTJE narrows by decreasing  $R_L$, since the load curves cease to intercept the SQUID $IV$ characteristics even at the lowest values of $P_{in}$. Indeed, the engine is not able to sustain  power if the load resistor is too low (we measured $R=0.5$ M$\Omega$ for the present device, see the inset of Fig. \ref{Fig3}f). 
The phase tunability of the thermoelectric effect allows to control power production of the BTJE with $\Phi$ (see Fig. \ref{Fig3}g), as theoretically predicted for a similar Josephson interferometer \cite{Marchegiani2020_3}.
The value of $P_L$ can be controlled over a larger range of amplitudes for smaller resistive loads at the cost of
power generation occurring in a narrower flux window. 
By contrast, large loads lead to reduced power tunability although occurring over a larger operation range in $\Phi$.

We now discuss the impact of bath temperature ($T$) on the BTJE performance. Figure \ref{Fig4}a shows the dependence of $P_L$ on $P_{in}$ measured at different temperatures for $R_L=2$
M$\Omega$, and $\Phi=0.33\Phi_0$. By rising $T$, the output power generally  decreases, and the BTJE operation range in $P_{in}$ narrows, since the bipolar thermoelectric effect weakens at higher temperatures. 
Yet, the BTJE operates up to $250$ mK, corresponding to $\sim 40\%$ of the critical temperature of S$_2$.
The expected non-monotonic dependence of $P_L$ on the input power can be appreciated at $T=250$ mK. This behavior can be ascribed to a sizable increase of the electron-phonon coupling in S$_1$ by rising the bath temperature \cite{Giazotto2006,Timofeev2009} and, thereby, to the resulting difference in the $P_{in}$ dependence of the thermal gradient established across the BTJE joined to the nonlinearity intrinsic to bipolar thermoelectricity.
Moreover, the $\Phi$-tunability of the output power turns out to be more efficient at higher values of bath temperature, as shown in Fig. \ref{Fig4}b. 
Indeed, a weaker thermoelectric effect is more sensitive to the magnetic flux (see also Fig. \ref{Fig2}e). Therefore, depending on the specific application, the operation temperature of the BTJE can be chosen in order to maximise the output power (i.e., at low bath temperature) or the flux sensitivity (i.e., at high bath temperature). 

\subsubsection*{\textbf{Conclusions}}
Our work presents a fully-bipolar thermoelectric effect occurring in superconducting tunnel junctions, thereby unveiling the possibility for significant thermoelectricity also in PH symmetric systems \cite{Marchegiani2020}. Differently from conventional linear thermoelectric effects \cite{Ashcroft1976, Abrikosov1988}, when subject to a large thermal bias, our Josephson junctions show a remarkable \emph{bipolar} power generation, which stems from a non equilibrium-induced \emph{spontaneous} PH symmetry breaking. 
Indeed, the sign of the generated voltage depends on the bias history of the system for a given thermal gradientimposed across the structure only, since the electronic configuration cannot be changed in metallic superconductors.
In particular, our superconducting junctions yield thermovoltages up to about $\pm150\;\mu$V corresponding to a nonlinear Seebeck coefficient $\mathcal{S}\simeq \pm300\;\mu$V/K. Strikingly, this value is  $\sim 10^5$ times larger than the linear Seebeck coefficient of a normal metal, and is comparable to that achievable in a quantum dot \cite{Mani2009,Prete2019} operating at the same temperature. 
The integration of thermally-biased superconducting tunnel junctions in a properly designed Josephson interferometer \cite{Kemppinen2008,Fornieri2016} allows fine control of the thermoelectric effect via an external magnetic flux \cite{Marchegiani2020_3}, owing to the strong competition between the Josephson coupling and thermoelectricity. 
We then exploit the resulting bipolar thermoelectric Josephson engine \cite{Marchegiani2020_2} to power a generic load resistor kept at room temperature. In this configuration, a direct current injection has the double role of igniting the engine, and selecting its output polarity. 
The BTJE delivers up to $\sim 24$ fW corresponding to an areal output power  density of $\sim140$ nW/mm$^2$ or, equivalently, to a power per conductance unit of $\sim 190$ pW/S. 
Therefore, the thermoelectric power could be naturally scaled in BTJE by increasing the junction size or through a multiple parallel configuration of several devices. 
In addition, the circuit controlling the BTJE defines a hysteretic $IV$ characteristic, which can be exploited to realize a potentially-fast volatile thermoelectric memory cell written or erased by a bias current \cite{Patent}.

In terms of sizable-power thermoelectric production, an on-chip generator could consist of an array of several BTJEs connected in series or parallel, as shown in Fig. \ref{Fig4}c. The series connection can advantageously be realized only by exploiting the described \emph{bipolar} device, since otherwise the voltage drop across each element would not simply add. Indeed, in conventional unipolar thermoelectric devices this configuration can only work by using a sequence of \emph{n} and \emph{p} materials. 
Here, the same junction takes spontaneously the role of the two elements \cite{Marchegiani2020_2}.
The total output power is $P_{tot}=nP$  in both configurations (with $n$ the number of elements), and while the series connection ensures high voltage production ($V_{tot}=nV$), the parallel connection guarantees high current generation ($I_{tot}=nI$). 
We also note that the BTJE already exploits the parallel connection of three thermoelectric elements (i.e., the S$_1$IS$_2$ junctions) to increase the total output current (see Fig. \ref{Fig1}b). 

From the side  of possible applications, the BTJE might find direct utilization  in superconducting quantum technology \cite{Ladd2010,Siddiqi2021,Polini2022} through the implementation of engines, power generators, electronic devices \cite{Braginski2019}, memories, radiation sensors \cite{Tero2018} and switches. Yet, alternative realizations of the BTJE might exploit extended Josephson tunnel junctions immersed in a parallel magnetic field \cite{Martinez2014}, or subject to spin-filtering in order to suppress the Josephson coupling \cite{Ozaeta2014,Machon2013,Bergeret2018,Linder,Kolenda2016}. 
Finally, the above bipolar thermoelectric effect is expected to occur in several physical systems characterized by intrinsic PH symmetry in the presence of a large temperature gradient (nonlinear response regime): thermoelectricity, indeed, requires only tunnel junctions where the hot and cold electrodes possess a \textit{gapped} and \textit{monotonically-decreasing} density of states, respectively \cite{Marchegiani2020}. Our study is then pivotal for groundbreaking investigations of nonlinear thermoelectric effects in different systems ranging from semiconductors and low-dimensional electronic materials to high-temperature superconductors and topological insulators.

\section*{Acknowledgements}
The authors wish to thank for useful discussion Dr. T. Novotny, Dr. K. Michaeli, Prof. L. Amico, and Prof. F. Strocchi. 
We acknowledge the European Research Council under Grant Agreement No. 899315-TERASEC, and  the  EU’s  Horizon 2020 research and innovation program under Grant Agreement No. 800923 (SUPERTED) 
and No. 964398 (SUPERGATE)
for partial financial support.  A.B. acknowledges the SNS-WIS joint lab QUANTRA, funded by the Italian Ministry of Foreign Affairs and International Cooperation and the Royal Society through the International Exchanges between the UK and Italy (Grants No. IEC R2 192166 and IEC R2 212041)

\section*{Author contributions} \label{sec:Author contributions}
F.P. fabricated the devices. 
G.G. and F.P. performed the experiments, and analysed the data with inputs from F.G..
G.M. and A.B. developed the theoretical model describing the experiment.
All the authors wrote the manuscript.
F.P. and F.G. conceived the experiment.
F.G. supervised and coordinated the project. All authors discussed the results and their implications equally at all stages. 

\section*{Additional information} \label{sec:Additional information}
Supplementary Information is available for this paper.

Correspondence and requests for materials should be addressed to F.G..

The authors declare no competing interests.

\section*{Methods}
\subsection*{Devices fabrication}
The devices were nanofabricated by a single electron-beam lithography (EBL) step, three-angle shadow-mask metals deposition onto a Si wafer covered with 300nm-thick thermally-grown SiO$_2$ through a suspended bilayer resist mask, and in-situ metal oxidation to define the tunnel junctions. 
The evaporation and oxidation processes were performed in an ultra-high vacuum (UHV) electron-beam evaporator with a base pressure of $10^{-11}$ $\textrm{Torr}$. 
At first, the superconducting heaters were deposited by evaporating a 12-nm-thick aluminum film at an angle of $30^\circ$. Then, the film was exposed to 1 Torr of $\textrm O_{2}$ for 20 minutes to create the AlO$_x$ layer forming the tunnel barriers. Subsequently, the 14-nm-thick aluminum island (S$_1$) was evaporated at $0^\circ$ and oxidized in 1 Torr of pure oxygen atmosphere for 30 minutes to realize the tunnel junctions of the  SQUID. Finally, an aluminum/copper bilayer (S$_2$, $t_{Al}=14$ nm and  $t_{Cu}=11$ nm) was deposited at an angle of $-30^\circ$ to form the remaining arms of the interferometer.  

\subsection*{Measurement set-up}
All measurements were performed in a filtered He$^3$-He$^4$ dry dilution refrigerator (Triton 200, Oxford Instruments) at different bath temperatures ranging from 30 mK to 600 mK. The transport properties of the double-loop SQUID were recorded in a standard two-wire configuration by applying a voltage bias through a floating source (GS200, Yokogawa), and by measuring the current with a room-temperature current pre-amplifier (Model 1211, DL Instruments). The Joule heaters were energized by a battery-powered voltage source (SIM 928, Stanford Research Systems). Finally, the magnetic flux piercing the SQUID was provided by a superconducting solenoid driven by a low-noise current source (GS200, Yokogawa).

\subsection*{Device parameters}
The areas of the central, left and right junction are obtained from the SEM picture of the device, and get the value $A_0=8.5\times10^{-2}\;\mu$m$^2$, $A_A=4.5\times10^{-2}\;\mu$m$^2$ and 
$A_B=4\times10^{-2}\;\mu$m$^2$, respectively. The resulting total area of the junctions composing the interferometer is $A_{int}=A_0+A_A+A_B=1.7\times10^{-1}\;\mu$m$^2$ (in good agreement with the SQUID interference patterns).
The normal-state tunnel resistance of all tunnel junctions is obtained from the experimental $IV$ characteristics. 
The tunnel resistance of the double-loop interferometer results from the parallel connection of its three junctions, thus it reads $R_{T}=\frac{R_0R_AR_B}{R_0R_A+R_0R_B+R_AR_B}\simeq9$ k$\Omega$, where $R_0$, $R_A$ and $R_B$ are the normal-state resistances of the central, left and right junction, respectively. 
The normal-state resistance of the two tunnel junction Joule heaters is $R_{h1}\simeq25.2$ k$\Omega$, and $R_{h2}\simeq26.6$ k$\Omega$, respectively. 
The shown two-wire characteristics include the resistance of the cryostat filters. 
Each line contributes to the transport with a resistance $R_{f}=1.1$ k$\Omega$.
The S$_1$ island is characterized by a zero-temperature energy gap $\Delta_{0,1}\simeq220\;\mu$eV, thus providing a superconducting coherence length $\xi_{S1}=\sqrt{\hbar D_{Al}/\Delta_{0,1}}\simeq82$ nm (where $\hbar$ is the reduced Planck constant, and $D_{Al}=2.25\times 10^{-3}$ m$^{2}$s$^{-1}$ is the diffusion constant of the Al film).
The Al/Cu bilayer forming S$_2$ shows a zero-temperature energy gap $\Delta_{0,2}\simeq80\;\mu$eV. S$_2$ lies within the Cooper limit, since $t_{Al}\ll \xi_{Al}\simeq82$ nm and $t_{Cu}\ll \xi_{Cu}=\sqrt{\hbar D_{Cu}/(2\pi k_B T)}\simeq127$ nm (where $D_{Cu}=8\times 10^{-3}$ m$^{2}$s$^{-1}$ is the diffusion constant of Cu, $k_B$ is the Boltzmann constant and $T=600$ mK is the highest operation temperature of the SQUID). Therefore, the superconducting coherence length of S$_2$ is $\xi_{S_2}=\sqrt{\hbar (t_{Al}D_{Al}+t_{Cu}D_{Cu})/[(t_{Al}+t_{Cu})\Delta_{0,2}]}\simeq190$ nm.
Finally, the ratio between the energy gaps of S$_2$ and S$_1$ is $r=\Delta_{0,2}/\Delta_{0,1}\simeq0.36$.

\subsection*{Quasiparticle current in the S$_1$IS$_2$ junction}
When the Josephson contribution is fully suppressed, by using a simple semiconductor model \cite{Tinkham} it is straightforward to express the quasiparticle current in S$_1$IS$_2$ tunnel junctions
as
\begin{equation}
I=\frac{1}{eR_T}\!\!\int_{-\infty}^{+\infty}\!\!\!\!\!\!\!\!\!\!d\epsilon N_{1}(\epsilon )N_{2}(\epsilon+eV)[f(\epsilon,T_1)-f(\epsilon+eV,T_2)],
\label{eq:iqp}
\end{equation}
where $R_T$ is the normal-state resistance of the junction, $e$ is the electron charge, $V$ is the voltage drop across the junction,
$N_i(\epsilon)$ and $f(\epsilon,T_i)$ (with $i=1,2$) are the quasiparticle density of states (DoSs) and the quasiparticle distribution functions, respectively. We work in the quasi-equilibrium regime \cite{Giazotto2006,Fornieri2017}, where the electronic temperature of each electrode is assumed to be well defined, and the occupation is expressed by the Fermi distribution function, $f(\epsilon,T_i)=[1+\exp(\epsilon/k_BT_i)]^{-1}$.
We assume a smeared BCS DoS for the superconducting $i$-lead, i.e., $N_i(\epsilon)=|\Re[(\epsilon+i\Gamma_i)/\sqrt{(\epsilon+i\Gamma_i)^2-\Delta_i(T_i)^2}]|$, where $\Delta_i(T_i)$ is the temperature-dependent energy gap, and $\Gamma_i$ is the Dynes parameter  accounting for quasiparticle states within the energy gap \cite{Dynes1984}. 
The PH symmetry of the superconducting leads is reflected in the symmetry of the DoSs with respect to the energy, i.e.,  $N_i(\epsilon)=N_i(-\epsilon)$. 
By using this symmetry, it is straightforward to demonstrate the anti-symmetry of the tunnelling current, $I(-V)=-I(V)$. 
The essential features of the thermoelectric effect are well captured by this modeling \cite{Marchegiani2020}, and the theoretical curves shown in Fig. \ref{Fig2}a and b are obtained by fitting the experimental data with Eq. \eqref{eq:iqp} (see SI for details). 
For two different superconductors with $\Delta_{0,1}>\Delta_{0,2}$ kept at the same temperature (i.e., for $T=T_1=T_2$), the above expression predicts a purely dissipative behaviour. 
In the subgap regime [i.e., for $|eV|<(\Delta_{1}(T_1)+\Delta_{2}(T_2))$], the current is suppressed, and displays a matching peak at $V_{p}=\pm(\Delta_{1}(T_1)-\Delta_{2}(T_2))/e$ \cite{Shapiro1962}, whereas a sharp transition to the normal-state occurs at voltages $\pm(\Delta_{1}(T_1)+\Delta_{2}(T_2))/e$. 
In the presence of a temperature gradient or non-equilibrium, Eq. \eqref{eq:iqp} admits also an \textit{absolute negative} conductance (ANC) \cite{Spivak,Ger1}, meaning that the current can flow against the voltage bias (i.e., $IV<0$) hence acting as a thermoelectric generator. 
In particular, one can show that the S$_1$IS$_2$ system can generate thermoelectric power for $T_1>T_2$ when the thermal gradient existing across the junction  is sufficiently large ($T_1\gtrsim T_2\Delta_{0,1}/\Delta_{0,2}$) \cite{Marchegiani2020,Marchegiani2020_2}. 
Furthermore, the antisymmetry of the $IV$ characteristic with voltage  implies the peculiar \emph{bipolarity} of this thermoelectric effect. Notably, quantum dots can generate thermo-currents flowing in both directions for a given temperature gradient, but a change of their electronic configuration by local electrostatic gating is required \cite{Josefsson2018}. Instead, the spontaneous breaking of the PH symmetry guarantees one of the two solutions without the need of any external intervention in our devices.
The S$_1$IS$_2$ system shows optimal performance when the gap ratio fulfills the condition $\Delta_{0,2}/\Delta_{0,1}=0.2 - 0.5$. In addition, an increase of the temperature gradient is not necessarily beneficial for thermoelectricity, since this effect is strongly nonlinear. When the system is thermoelectric, there are, at least, two finite Seebeck voltages. We observe that these  Seebeck voltages are slightly larger than the value of the matching peak, i.e., $(\Delta_{0,1}-\Delta_{0,2})/e\lesssim |V_{th}^\pm|$ \cite{Marchegiani2020}. The maximum Seebeck voltages are mainly determined by the gap values, and they are only weakly-dependent on the thermal gradient occurring across the junction. 

\subsection*{Seebeck coefficient}
The Seebeck coefficient of a normal metal in the presence of energy-dependent scattering mechanisms (diffusive limit) can be calculated through the Mott-Jones equation \cite{Mott1958},
 $\mathcal{S}_{linear}\simeq \left(\frac{\pi^2}{3}\right)\left(\frac{k_B}{e}\right)\left(\frac{k_BT}{E_{F,0}}\right)x$,
where $E_{F,0}$ is the zero-temperature Fermi energy of the metal and $x=2.78$ is a numerical constant that depends on the energy dependences of various charge transport
parameters in Al. The resulting Seebeck coefficient of Al takes the values $S_{Al}(T=300\;\text{K})\simeq-1.74\;\mu$V/K and $S_{Al} (T=650\;\text{mK})\simeq-3.8$ nV/K, where $E_{F,0}\simeq11.6$ eV. 
Since Al is typically superconducting at subKelvin temperatures, we can only theoretically estimate the Seebeck coefficient. Notably, experiments investigating Al at temperatures higher than its critical temperature reported values for the Seebeck coefficient of the same order of magnitude as the above theoretical estimate \cite{vanHarlingen1984}.

\section*{Data availability}
All other data that support the plots within this paper and other findings of this study are available from the corresponding authors upon reasonable request.

\definecolor{goodgreen}{rgb}{0.1,0.5,0}

\hyphenation{ALPGEN}
\hyphenation{EVTGEN}
\hyphenation{PYTHIA}

\renewcommand{\thefigure}{S\arabic{figure}}
\renewcommand{\theequation}{S\arabic{equation}}

\section*{Supplementary Information for \\Bipolar Thermoelectric Josephson Engine}

\section{Model of the double-loop interferometer}
The interference pattern of a double-loop interferometer based on three tunnel Josephson junctions can be described by assuming a sinusoidal current-to-phase relation (CPR) for each junction and fluxoid quantization for the two loops. Therefore, the critical current of a double-loop interferometer can be written as \cite{Kemppinen2008,Ronzani2014,Fornieri2016}

\begin{equation}
     I_C(\Phi)=\max_{{\varphi_0}}\bigg{[} I_0\sin{(\varphi_0)}+ I_A\sin{\left(\varphi_0+2\pi\frac{\Phi_A}{\Phi_0}\right)} 
     + I_B\sin{\left(\varphi_0-2\pi\frac{\Phi_B}{\Phi_0}\right)}\bigg{]},
     \label{EqSQUID}
\end{equation}
where $\varphi_0$ and $I_0$ are the phase drop and the critical current of the central junction, $I_A$ is the critical current of the left junction, $I_B$ is the critical current of the right junction, $\Phi_0$ is the magnetic flux quantum, $\Phi_A$ is the magnetic flux piercing the left ring and $\Phi_B$ is the magnetic flux piercing the right ring. The total magnetic flux piercing the interferometer is $\Phi=\Phi_A+\Phi_B$. In order to take into account the asymmetry of the loop areas, we introduce the asymmetry coefficient $\alpha$ providing $\Phi_A=[(1+\alpha)\Phi]/2$ and $\Phi_B=[(1-\alpha)\Phi]/2$. By using the experimental values of $I_C(\Phi=0)$ recorded at each temperature, we fit the interference patterns presented in Fig. 1e of the main text, thus estimating $\alpha=0.003$, $r_A=I_A/I_0=0.63$ and $r_B=I_B/I_0=0.65$. We note that our device satisfies $|r_A-r_B|\leq1$ and $r_A+r_B\geq1$, therefore allowing to obtain perfect critical current suppression at specific values of $\Phi_A$ and $\Phi_B$. Indeed, we measure a supercurrent suppression of about $1.75$\textperthousand.

\section{Estimate of the electronic temperature}
The bipolar thermoelectric Josephson engine requires a finite temperature gradient to realize the thermoelectric effect. In the experiment, we do not directly measure the electronic temperatures of S$_1$ and S$_2$ ($T_1$ and $T_2$, respectively).
The theoretical curves displayed in Fig. 2a-b of the main text (dashed lines) are obtained through a two-parameter ($T_1$, $T_2$) fitting procedure performed on the current versus voltage characteristics under power injection. 
As we detail below, the main junctions parameters, such as the zero-temperature gaps of the two superconductors ($\Delta_{0,1},\Delta_{0,2}$), are precisely estimated through a preliminary investigation of the equilibrium charge transport. 
In order to highlight the nature of the thermoelectric effect, we rely on a minimal model: we consider the quasiparticle current and replace the three junctions of the interferometer by a single junction with effective normal-state resistance $R_T$ (the measured value of the normal-state resistance of the whole interferometer). 
The overall quality of the fits demonstrates the presence of the thermal gradient, and the robustness of the thermoelectric conversion with respect to unavoidable non-idealities present in the structure. 
The details of the model for quasiparticle transport in our systems are provided in the Methods section of the main text.

\subsection{Equilibrium analysis}

\begin{figure}[t!]
\centering
\includegraphics[width=\linewidth]{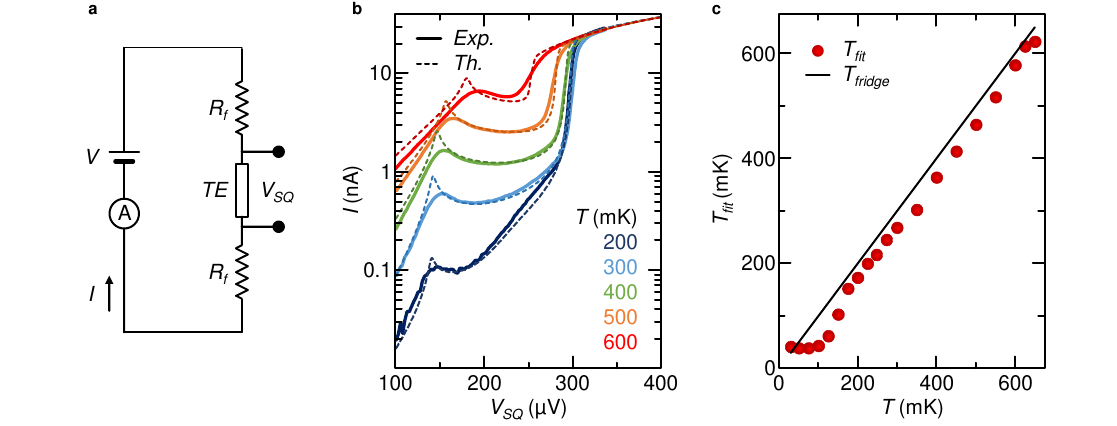}
\caption{\textbf{Equilibrium analysis of the current-voltage characteristics.} 
\textbf{a} Schematic electronic circuit for the measurement of the bipolar thermoelectric effect, where the thermoelectric element (TE) is in series with two $RC$ filters present on each line of the dilution refrigerator. Here, $V_{SQ}$ represents the voltage drop across the thermoelectric element (TE), $I$ is the current, and $R_f=1.1$ k$\Omega$ is the resistance of each filter. 
\textbf{b} Experimental $I(V_{SQ})$ characteristics (solid lines) plotted together with the fitted curves (dashed lines) for different values of bath temperature. \textbf{c} Temperature obtained from the fitting of the $I(V_{SQ})$ characteristics ($T_{fit}$, dots) and fridge temperature ($T_{fridge}$, line) for several measurements.}
\label{figS0}
\end{figure}

We measured the current-voltage tunnel characteristics $I(V)$ of the device for a zero input power ($P_{in}=0$) at a magnetic flux $\Phi=0.33\Phi_0$ (so to minimize the Josephson coupling). The measurement scheme is shown in Fig. \ref{figS0}a.
As a first step, we observe that the voltage drop across the SQUID ($V_{SQ}$) is different from the measured value $V$, due to the presence of $RC$ filters on the cryostat measurement lines. In particular, the SQUID voltage reads $V_{SQ}=V-2R_f I(V)$, where $R_f=1.1$ k$\Omega$ is the single filter resistance. 
The measured curves are shown in Fig. \ref{figS0}b in logarithmic scale (solid lines) for different values of bath temperature ($T$) in the range [30, 650] mK. 
The current is clearly non-monotonic in the voltage bias ($V_{SQ}$). The characteristics display a temperature-dependent peak structure for $V_{SQ}\sim V_p$, the "matching peak" in the jargon of the superconducting community, and an Ohmic behavior $I\sim V_{SQ}/R_{T}$ (with $R_{T}\simeq 9$ k$\Omega$ the total normal-state resistance of the SQUID) after a sharp current increase for $V_{SQ}\sim V_{max}$.
In our modeling of the S$_1$IS$_2$ junction (see Eq. 1 of the main text), the values $V_p\sim[\Delta_{1}(T_1)-\Delta_{2}(T_2)]/|e|,V_{max}\sim[\Delta_{1}(T_1)+\Delta_{2}(T_2)]/|e|$ (with $e$ the electron charge) yield information on the superconducting energy gap of S$_1$ and S$_2$.   
Therefore, we can immediately estimate the zero-temperature energy gaps $\Delta_{0,1}\sim220\;\mu$eV and $\Delta_{0,2}\sim 80\;\mu$eV, from the lowest temperature $IV_{SQ}$ curves.

We make an \emph{ansatz} for the temperature dependence of the gaps of S$_1$, and S$_2$, i.e., $\Delta_i(T)=\Delta_{0,i}\delta(T/T_{C,i})$ with $i=1,2$. 
Here, $\delta(T/T_{C,i})$  expresses the universal dependence \cite{Tinkham} of the superconducting energy gap (in units of the zero-temperature value) as a function of the reduced temperature in the BCS weak-coupling limit. 
In particular, an approximate form of this numerical function is given by the expression $\delta(x)=\tanh({1.74\sqrt{x^{-1} -1}})$, with a maximum deviation of the order of $2\%$.
The critical temperatures of S$_1$ and $S_2$ were estimated experimentally as $T_{C,1}\sim1.4$ K, and $T_{C,2}\sim625$ mK, respectively.
Notably, the S$_1$ gap-to-critical current ratio, $\Delta_{0,1}/(k_BT_{C,1})$, is very close to the BCS prediction, while $\Delta_{0,2}/(k_BT_{C,2})$ for the bilayer (S$_2$) deviates of about the 20\%. This deviation may be ascribed to the spatial variation of the bilayer yielding a different gap amplitude at the three junctions (here treated as constant), or to the inverse proximity effect \cite{deGennes} which is used to properly engineer the gap, i.e., the film is a composite artificial superconductor.
Finally, we need to establish the values of the Dynes parameters \cite{Dynes1984} in the smeared DoS of S$_1$ and S$_2$. In our calculations, we set $\Gamma_{1}=2\times 10^{-3}\Delta_{0,1}$, $\Gamma_{2}=2\times 10^{-2}\Delta_{0,2}$, which guarantee a good matching between the $IV_{SQ}$ experimental characteristics and the theoretical predictions, as we will see in the following. 

In order to test the validity of the S$_1$IS$_2$ modeling and the parameters, we performed a one-parameter ($T_{\rm fit}$) fit on the experimental curves, assuming thermal equilibrium $T_1=T_2=T_{\rm fit}$. In the fitting procedure, we excluded the low-bias values ($V_{SQ}\to 0$), where the contribution of the residual Josephson effect can not be neglected. 
The $IV_{SQ}$ characteristics fitted curves are shown in Fig. \ref{figS0}b (dashed lines) for different values of bath temperature. 
While the theoretical curves reproduce the overall behavior somewhat satisfactorily, we observe that the S$_1$IS$_2$ modeling of the $IV_{SQ}$ characteristics leads to a systematic overestimate of the height of the matching peaks. 
This deviation is likely to be associated to our simplified modeling. Indeed, we consider a single junction instead of three (which may determine some inhomogeneity of the parameters among the different junctions) introducing a possible effective smearing parameter of the peaks which, for simplicity, we did not account here. Alternatively, this smearing in S$_2$ may be associated to the presence of the normal-metal Cu layer used to engineer the artificial superconductor. 
 
Figure \ref{figS0}c displays the temperature ($T_{fit}$, points) obtained from the fit of the $IV_{SQ}$ curves taken at different  bath temperatures ($T$). 
The line $T_{fit}=T_{fridge}$ is also indicated for a comparison (solid line). The temperature profile is consistent with the fridge temperature with a good degree of accuracy. 
Significant deviations appear at lower temperatures ($T\lesssim 175$ mK), where the signal becomes small (due to the exponential dependence of the $I(V_{SQ})$ characteristic on temperature of the electrodes) and more noisy, thus making a reliable fitting not possible. 
 In summary, the equilibrium analysis 
demonstrates the possibility of extracting the electronic temperatures of S$_1$, S$_2$ with good accuracy by fitting the $IV_{SQ}$ characteristics. 

\subsection{Out-of-equilibrium Calibration}
The values of the energy gaps and the Dynes parameters of the two superconductors (S$_1$ and S$_2$) forming the junctions were extracted from the analysis of the $IV_{SQ}$ characteristics at thermal equilibrium, as discussed above. 
In this section, we fix these parameters ($\Delta_{0,1}\sim220\;\mu$eV, $\Delta_{0,2}\sim80\;\mu$eV, $\Gamma_{1}=2\times 10^{-3}\Delta_{0,1}$ and $\Gamma_{2}=2\times 10^{-2}\Delta_{0,2}$) and determine the temperatures $T_1$ and $T_2$ for values of the heating power ($P_{\rm in}$) used in our experiments. 
To validate our analysis, we discuss two independent procedures: i) calibration of the hot temperature $T_1$ based on the $P_{in}$ evolution of the gap threshold; ii) two-parameter ($T_1$ and $T_2$) fitting of the complete $IV_{SQ}$ characteristics for different values of $P_{in}$.


\subsubsection{Experimental method to create a thermal gradient across the structure}
The temperature gradient across the $S_1IS_2$ junctions is created by increasing the temperature of $S_1$ ($T_1$) through the injection of power $P_{in}$ directly in $S_1$. The latter is generated by voltage biasing ($V_h$) the series connection of two superconducting electrodes tunnel-coupled to $S_1$ \cite{Giazotto2006}. As a consequence, $P_{in}= V_{h}^2/2R_h$, where $R_h=51.8$ k$\Omega$ is the normal-state series resistance of the two heaters, since the Joule heating is dissipated on the two sides of the junction. Therefore, the minimum value of $P_{in}$ is limited by the minimum voltage bias $V_h\geq 2[\Delta_1(T_1)]+\Delta_h(T_h)]/|e|$, where $\Delta_h$ is the temperature-dependent energy gap of the superconductor composing the heaters, and $T_h$ their electronic temperature. Hence, the minimum heating power that can be injected in $S_1$ at a bath temperature $T=30$ mK is $P_{in}\geq 10$ pW.
Moreover, the main source of electron thermalisation in $S_1$ ($P_{e-ph,S1}$) is provided by electron-phonon interaction with the lattice. 
In a superconducting thin film, the electron-phonon coupling turns out to be exponentially suppressed at low temperature owing to the presence of the energy gap in the density of states \cite{Giazotto2006}. In the above conditions, even a somewhat limited amount of injected power can lead to a substantial enhancement of the quasiparticle temperature in a superconductor. For the BTJE, we have $P_{in} \gg P_{e-ph,S1}$ at $T=30$ mK and, thereby $T_1\gg T_2 \geq T$, where $T_2$ is the electron temperature of $S_2$.

\subsubsection{Gap-sum threshold calibration}
\begin{figure}[t]
\centering
\includegraphics[width=1\linewidth]{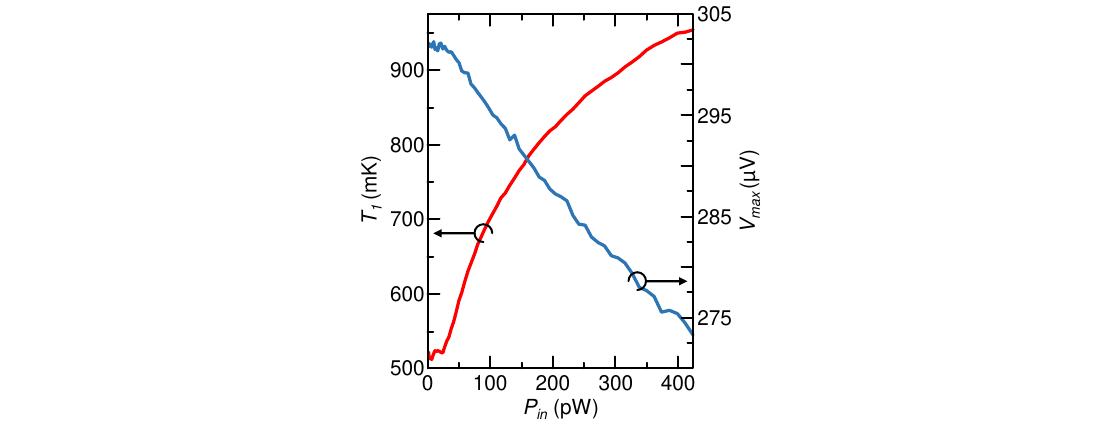}
\caption{\textbf{Gap-sum threshold calibration.} Blue curve: evolution of the voltage corresponding to the maximum of the differential conductance ($V_{max}$) as a function of the input power ($P_{in}$). Red curve: temperature $T_1$ of the hot superconductor (S$_1$) vs $P_{in}$  obtained through a numerical solution of Eq. \ref{eq:gamSupNE}.}
\label{figS1}
\end{figure}

We start with the gap-threshold estimate of the temperature $T_1$ of the hot electrode ($S_1$). 
As discussed above, the gap-threshold bias voltage in the S$_1$IS$_2$ modeling is associated to the gaps sum, i.e., $V_{max}\sim [\Delta_1(T_1)+\Delta_2(T_2)]/|e|$. This quantity is extracted from the experimental curves as the maximum of the differential conductance curves, $dI/dV_{SQ}$. 
Figure \ref{figS1} displays $V_{max}$ as a function of the $P_{in}$. As expected, $V_{max}$ decreases monotonically with $P_{in}$, since the energy gap $\Delta_1(T_1)$ is reduced due to the temperature increase.
Here, we assume S$_2$ to be well thermalized at the lowest bath temperature so that its gap is well approximated by the zero-temperature value, $\Delta_2(T_2)\approx\Delta_{0,2}$. Within this approximation, the temperature $T_1$ of S$_1$ is obtained by solving the following equation:  
\begin{equation}
|e| V_{\rm max}=\Delta_{0,1}\delta(T_1/T_{c,1})+\Delta_{0,2}.
\label{eq:gamSupNE}
\end{equation}
The solution of Eq. \ref{eq:gamSupNE} is displayed in  
Fig.~\ref{figS1} (solid red line). After a flat behaviour, $T_1(P_{in})$ increases monotonically with the input power for $P_{in}= 22.5$ pW.
We note that the following calibration is not reliable for small values of $P_{in}$. 
As a matter of fact, the temperature evolution of  $\Delta_1(T_1)$ is exponentially suppressed for $T_1\lesssim 0.4 T_{C,1}$ so that it is fairly difficult to infer the correct hot temperature for those regimes giving the flat and noisy behaviour occurring for $P_{in}<$ 22.5 pW. 

\subsubsection{Complete current-voltage characteristic calibration}
We proceed now with a more refined calibration through numerical fitting of the complete current-voltage characteristic of the device. In the Methods section of the main text we provided the expression of the tunnelling current of the quasiparticle transport (Eq. 1 of the main text). 
Since we know the temperature dependence of the gaps (as discussed in the equilibrium analysis) and the Dynes parameters of $S_1$ and $S_2$, the only two unknown quantities to fit the $IV_{SQ}$ curves are  the temperatures of the leads ($T_1$ and $T_2$). 

\begin{figure}
\centering
\includegraphics[width=1\linewidth]{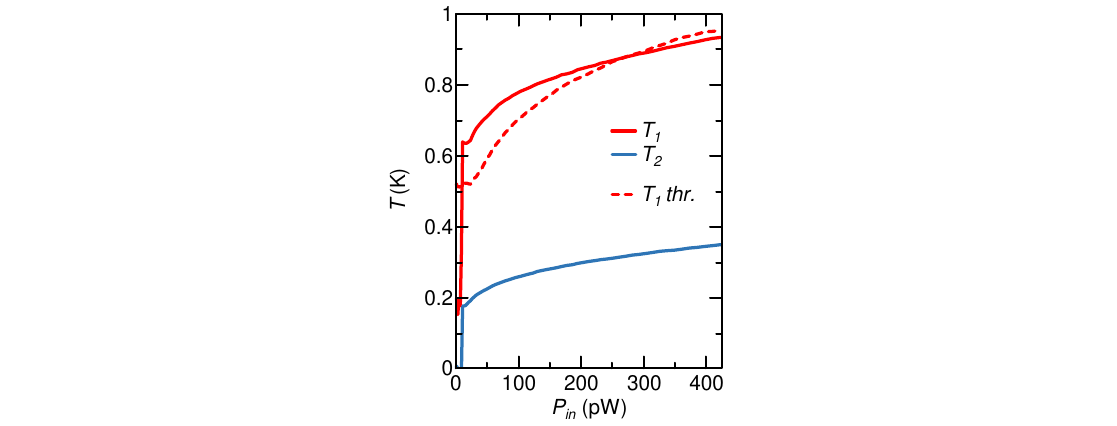}
\caption{\textbf{Complete current-voltage characteristic calibration.} Solid: temperature of the hot ($T_1$, red) and the cold superconductors ($T_2$, blue) as a function of the input power $P_{in}$. Dashed: the gap-sum calibration for the temperature of $T_1$ is displayed for a comparison.}
\label{figS2}
\end{figure}

The results of this fitting are shown in Fig.~\ref{figS2}, where the temperatures $T_1$ and $T_2$ are plotted versus the input heating bias $P_{in}$ (solid curves). The fitting procedure fails at very low injecting power, i.e., for  $P_{in}< 10$ pW, where the $I(V_{SQ},T_1,T_2)$ characteristic depends very weakly on $T_1$ and $T_2$. We stress that the system is not thermoelectric for these low input powers. More precisely, the corresponding temperature gradient is insufficient to develop the nonlinear thermoelectricity in our device. As a result, the $I(V_{SQ})$ sensitivity with respect to the temperature difference is quite low.
Instead, the temperature characterization (calibration) is significantly enhanced in the presence of thermoelectricity, which is signaled by the condition $IV_{SQ}<0$. In such cases, the fitting returns the best values for the temperatures of the two leads. The temperature of $S_1$ increases monotonically with the input power from a minimum temperature $T_1(P_{in}=22.5\; {\rm pW})\sim 0.65$ K to the value $T_1(P_{in}=422.4\; {\rm pW})\sim 0.93$ K for the maximum power considered.
Notably, the overall calibration is in good agreement with the previous estimation [$T_1(P_{in})$, dashed line] with a maximum difference of about $100$ mK. 

Moreover, the fitting procedure gives information about the temperature of the bilayer $T_2(P_{in})$. Interestingly, $T_2$  increases monotonically with $P_{in}$ too (solid blue line). This behavior is associated with the heat-current flowing through the three junctions of the interferometer. This contribution represents the heat current transferred from the hot ($S_1$) to the cold ($S_2$) reservoir, as occurs in the operation of any standard thermodynamical engine. This interpretation could be further supported by imposing heat balance equations to the different parts of the system. Unfortunately, the complex geometry and the bilayer nature $S_2$, combined with the unknown thermal coupling with the substrate, makes this procedure unfeasible. 
Notably, the maximum temperature of $S_2$ is about $0.35$ K, where the gap is only slightly reduced with respect to the zero-temperature value $\Delta_2(T_2=0.35 {\rm K})\sim 0.93\Delta_{0,2}$, thus justifying \textit{a posteriori} the simplified approach that we followed in the previous section.  

In conclusion, the presented calibration methods clearly demonstrate that there is a clear temperature gradient across the interferometer. In the curves (theoretical and experimental) for the nonlinear Seebeck coefficient ($\mathcal{S}$) displayed in Fig. 2c of the main text, we considered the two-parameter fitting calibration. In the previous analysis, we took conservative  choices that may result in an overestimation  of the thermal gradient in the junction. Thus, the estimated nonlinear Seebeck coefficient is possibly slightly underestimated, since $\mathcal{S}$ is inversely proportional to the thermal gradient. Therefore, the real system may eventually even provide larger values of $\mathcal{S}$ than the estimations reported in the main text. 

\section{Hysteretic transport in the engine configuration}
\label{Hyst}
\begin{figure*}[t!]
    \centering
    \includegraphics[width=1\linewidth]{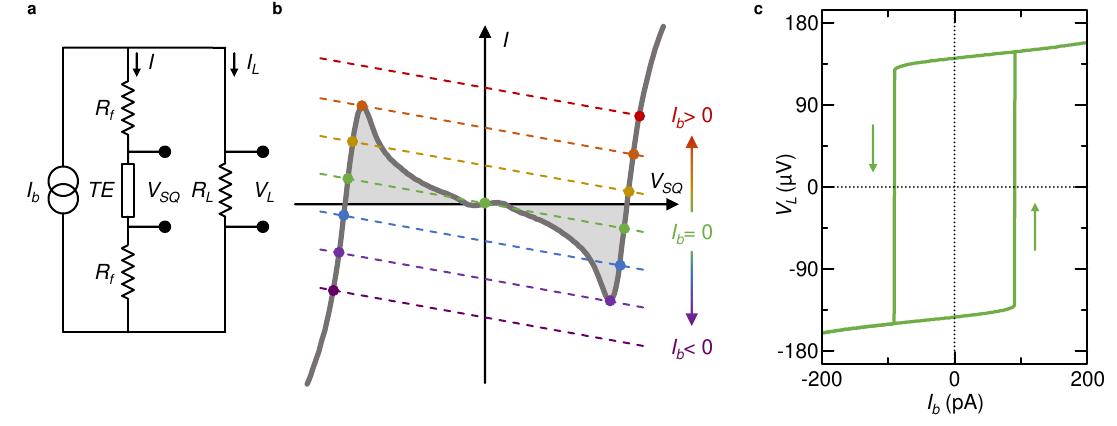}
    \caption{\textbf{Power generation in the engine configuration.}
    \textbf{a} Schematic electronic circuit implementing the heat engine. The parallel between the load resistor ($R_L$) and the series of thermoelectric element (TE) and two filters ($R_f$) is biased by a dc current generator ($I_b$). $V_L$ is the voltage measured across the load, while $V_{SQ}$ is the voltage generated by the thermoelectric element. 
    \textbf{b} Sketch of the current-voltage characteristic ($IV_{SQ}$, grey solid line) of the thermoelectric element intersected by a load-line for different positive and negative values of $I_b$. 
    \textbf{c} Theoretical hysteresis curve obtained by solving Eq. \ref{Eq:I(V)} for $R_L=2$ M$\Omega$ with the experimental $IV_{SQ}$ of the thermoelectric element recorded at $P_{in}=10$ pW.}
    \label{Fig:CurveIsteresi}
\end{figure*}

The electronic transport of the circuit implementing the bipolar thermoelectric Josephson engine (BTJE) shows a hysteretic behavior with the external bias current, as shown in Fig. 3c-d of the main text. The electronic circuit used to operate the BTJE is schematized in Fig. \ref{Fig:CurveIsteresi}a. Here, the thermoelectric element is connected in series to the two low-pass filters placed in the lines of the cryostat ($R_f=1.1$ k$\Omega$). 
We note that $I$ and $V_{SQ}$ are the current and the voltage provided by the thermoelectric element, respectively. 
Furthermore, an external current generator ($I_b$) powers the parallel connection of the thermoelectric element and the load resistance ($R_L$). The charge transport in the circuit is described by the following system of relations:
\begin{equation}
\label{Eq:sistema}
\begin{cases}
    I_{b}=I_{L}+I & \quad \quad \text{(a)}\\
    I_{L}=\frac{V_L}{R_L} & \quad \quad \text{(b)}\\
    V_L=2IR_f+V_{SQ} & \quad \quad \text{(c)} \\
\end{cases}
\end{equation}
where $I_L$ is the current flowing through the load resistor and $V_L$ is the associated voltage drop.

By substituting the Eq. \ref{Eq:sistema}a and b in Eq. \ref{Eq:sistema}c, we obtain the relation describing the direct-current steady state behavior of the circuit:
\begin{equation}
\begin{aligned}
I&= -I_L + I_b\\
&=-\frac{V_L}{R_L}+ I_b\\
&=-\frac{2IR_f+V_{SQ}}{R_L}+ I_b\\
&=-\frac{V_{SQ}}{R_L+2R_f}+\frac{R_L}{R_L+2R_f}I_b.\\
\label{Eq:I(V)}
\end{aligned}
\end{equation}

The result is a linear relation of $I(V_{SQ})$, where the load conductance ($\propto R_L^{-1}$) is the slope of the load-line and $I_b$ is the intercept with $I$-axis taking into account $R_L \gg R_f$. Note that, due to the non-monotonic behaviour of $I(V_{SQ})$, the previous equations can admit multiple solutions. In order to have a stable-solution, the differential conductance at the crossing must be positive, as discussed in Refs. \cite{Marchegiani2020,Marchegiani2020_2}.\\

Figure \ref{Fig:CurveIsteresi}b displays a sketch of the current-voltage characteristic of the thermoelectric element crossed by the load resistance at different values of the bias voltage (see Eq. \ref{Eq:I(V)}). By changing the bias current, the load-line sweeps the whole thermoelectric characteristic (dashed lines). The intersections between the two curves are highlighted with circles, which correspond to the solutions of Eq.(\ref{Eq:I(V)}) at different values of $I_b$. 
Thus, the theoretical curves describing the circuit in Fig. \ref{Fig:CurveIsteresi} can be directly obtained by substituting the measured $I(V_{SQ})$ characteristics in Eq. \ref{Eq:I(V)}. In this way, we can obtain the hysteretical curves measured experimentally (see Fig. 3c-d of the main text). In particular, Fig. \ref{Fig:CurveIsteresi}c shows the theoretical hysteresis curve obtained by solving Eq. \ref{Eq:I(V)} for $R_L=2$ M$\Omega$ with the experimental thermoelectric characteristics recorded at $P_{in}=10$ pW.
Notably, by using the measured $IV_{SQ}$ characteristic for the thermoelectric element, we can include also in the analysis the complex effects associated to the Josephson coupling. For example, a metastable state at $V\approx0$ can be observed in the presence of Josephson coupling. Intriguingly, this simple analysis of the load-line is also able to predict the ignition  of the thermoelectric elements. We will further investigate in  detail this rich phenomenology in the following  research.

\subsection{Hysteresis curves for different loads}
\begin{figure*}[t!]
    \centering
    \includegraphics[width=1\linewidth]{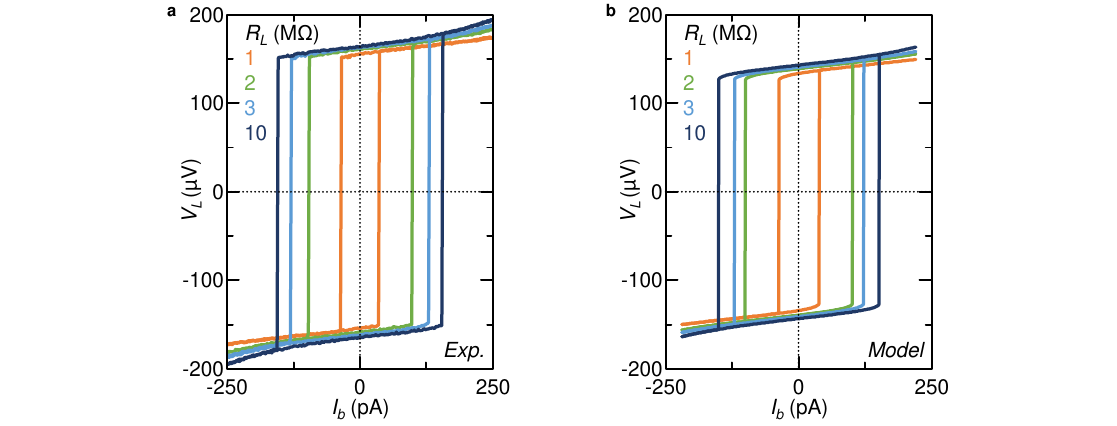}
    \caption{\textbf{Hysteresis for different load resistors.}
    Experimental \textbf{(a)} and theoretical \textbf{(b)} hysteretic $V_L(I_b)$ curves for different values of the load resistance ($R_L$). The width of the hysteresis cycle in the four curves increases with $R_L$. 
    }
    \label{FigS2}
\end{figure*}

Intriguingly, the analysis of the hysteresis in the $I_bV_L$ characteristics can be easily applied to any value of load resistance. 
The experimental hysteresis curves resulting for five different nominal values of the load resistance are shown in Fig. \ref{FigS2}a.
Since the slope of the load-line depends strongly on the value of the load resistance used ($R_L$), low values of $R_L$ do not cross the thermoelectric $IV_{SQ}$ characteristics.
This can be physically interpreted as the impossibility of the thermoelectric element to provide enough thermoelectrical power to the load. 
Indeed, this nonlinear thermoelectric effect necessarily produces a finite current to keep the spontaneous PH symmetry breaking, thus the element can clearly support a finite minimal load resistance. 
Differently, conventional linear thermoelectricity typically supports small loads by reducing the voltage and increasing the thermocurrent. 
The minimum value of the load resistance that can be supported by the junction is $R_{L,min}=V_p/I_p$ (with $\pm V_p$ and $\pm I_p$ the voltage and the current corresponding roughly to the thermoelectric peaks).
In our case, the minimum load powered by the engine is $R_{L,min}= 0.8$~M$\Omega$. 

We can observe that the width of the hysteresis increases with the load resistance, while the values of voltage generated at $I_b=0$ are almost independent from the load. 
The result is consistent with the previous discussion of Eq. \ref{Eq:I(V)}. Indeed, for a larger $R_L$, the slope of the load-line is smaller. Thus, a larger bias current is needed to invert the sign of the generated thermovoltage.  
Thus behavior is completely grabbed by the model describing the experimental circuit used to control the engine (see Fig. \ref{Fig:CurveIsteresi}a). Indeed, the theoretical $V_L(I_b)$ curves obtained by solving Eq. \ref{Eq:I(V)} at a fixed input power show the same dependence on $R_L$ (see Fig. \ref{FigS2}b for $P_{in}=10$ pW).

\section{Additional thermoelectric device}

\begin{figure} [t!]
    \centering
    \includegraphics[width=1 \linewidth]{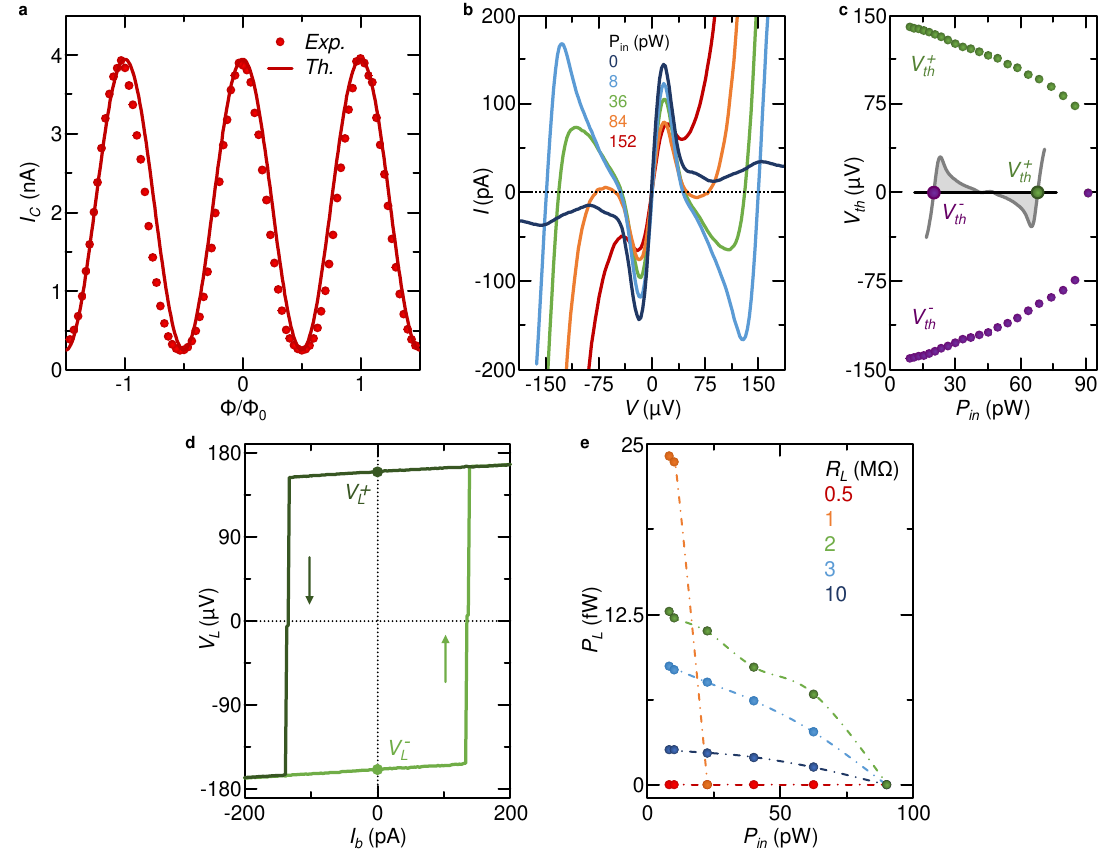}
    \caption{\textbf{Properties of the additional thermoelectric device.}
    \textbf{a} Experimental interference pattern of a double loop SQUID (dots). The pattern shows a regular sinusoidal shape, getting a suppression of about 5$\%$ with respect to the maximum critical current. A fitting curve is reported (red line) by employing the model for a double loop SQUID (see Eq. \ref{EqSQUID}). 
    \textbf{b} Current-Voltage characteristic in the presence of a thermal bias recorded at $\Phi=0.51\Phi_0$ and a bath temperature of 30 mK. The subgap $IV$ curves are registered for different values of the power injection ($P_{in}$). Bipolar thermoelectricity is present in a restricted range of the input thermal power, that is $P_{in}=8-90$ pW. 
    \textbf{c} Thermoelectric voltage ($V_{th}$) as a function of $P_{in}$ recorded at $\Phi=0.51\Phi_0$ and bath temperature of 30 mK. The voltage gets a bipolar maximum value $V_{th}\simeq \pm 141 ~\mu$V.
    \textbf{d} $V_L$ versus $I_b$ hysteresis curve recorded for $P_{in}=8$ pW and $R_L= 3$ M$\Omega$ at a bath temperature of 30 mK. $V_L^\pm$ are the voltage values generated by the heat engine at zero bias current ($I_b=0$). 
    \textbf{e} Engine output power generated on the load resistor ($P_L$) as a function of the injected power ($P_{in}$) recorded at a bath temperature of 30 mK for different values of $R_L$.}
    \label{Fig:SecDev_TE}
\end{figure}

To show the reproducibility of the effect, we tested and characterized also an additional device, which exhibits similar properties and behaviours.
The additional device presents a similar design of Fig.~(1c) of the main text. The details of the device fabrication and measurement set-up are quite the same of the sample used in the main text discussion and provided in the Methods section of the main text. 

The two loops of the additional device show almost the same area, thus $\Phi_A=\Phi_B$. Furthermore, the same critical current flows through the two lateral arms, while the central Josephson junction has double the critical current ($r_A \sim r_B \leq 0.5$). In this way, the second and the third term of Eq. \ref{EqSQUID} are less relevant and contribute by increasing the minimum critical current of the interferometer.
Figure \ref{Fig:SecDev_TE} shows the equilibrium interference pattern of the additional device recorded at a bath temperature of 30 mK.
The maximum critical current is $I_C \simeq 3.95$ nA, which is suppressed down to about 5$\%$. This value of suppression is about one order of magnitude smaller than the value reached in the device shown in the main text. The minimum critical current is $I_C=160$ pA recorded at $\Phi=0.51\Phi_0$ (with $\Phi_0$ the flux quantum).

Figure \ref{Fig:SecDev_TE}b shows the $IV$ of the device recorded at $\Phi=0.51\Phi_0$ and at a bath temperature of 30 mK for different values of input heating power ($P_{in}$). 
Differently from the device presented in the main text, the thermoelectric effect manifests in a smaller range of input power ($P_{in}= 8-90$ pW) due to the relevant presence of the Josephson current, which short-circuits the effect \cite{Meissner1927}. 
The maximum thermoelectric voltage is recorded at $P_{in}=8$ pW and gets the value $V_{th}=\pm141~\mu$V, as shown in Fig. \ref{Fig:SecDev_TE}c. The thermoelectric voltage decreases by rising $P_{in}$, because the superconducting gap of the heated island [$\Delta_1(T_1)$] is reduced by the local heating.
In addition, the gaps ratio [$\Delta_2(T_2)/\Delta_1(T_1)$] increases thus extinguishing the effect.
We note that despite the Josephson coupling is larger than in the device presented in the main text, the overall behavior of the $IV$ characteristics is qualitatively the same. Indeed, the Seebeck voltage has a very similar value and input power dependence even if in a smaller range of $P_{in}$.

We tested the additional device as BTJE by connecting the thermoelectric element in the circuit shown in Fig. 3a of the main text. The additional device shows the same qualitative behavior of the sample presented in the main text. 
Indeed, we obtained a hysteretic voltage-current characteristic, as shown in Fig. \ref{Fig:SecDev_TE}d for $R_L=$ 3 M$\Omega$ and $P_{in}=$ 8 pW at a bath temperature of 30 mK. The voltage across the load at zero bias current ($I_b=0$), produced by the thermoelectric element, takes the value $V_L^\pm=\pm 158~ \mu$V.
This value is very similar to the voltage of the matching peak [$(\Delta_{0,1}-\Delta_{0,2})/|e|$] and the voltage generated by the sample discussed in the main text. 
These voltages are given by the intersection between the load-line and the $IV_{SQ}$ characteristic of the thermoelectric element. The details of the circuit piloting the engine are practically the same of the main text. The details are discussed in Sec. \ref{Hyst} of the Supplementary Information. 
We note that the width of the hysteresis is slightly larger than that of the device presented in the main text (see Fig. \ref{Fig:SecDev_TE}d), since the maximum thermoelectric current produced by the additional device is higher. 

Figure \ref{Fig:SecDev_TE}e shows the output voltage power produced by the heat engine ($P_L$) as a function of $P_{in}$ for different values of $R_L$. At a given value of $P_{in}$, $P_L$ is larger for lower values of $R_L$ but the engine operates in a narrower window of input powers, as recorded for the device shown in the main text. 
For $R_L=0.5$ M$\Omega$ smaller than $R_{L,min}=$0.6 M$\Omega$, the engine does not produce any power on the load, as expected. Similarly to the other device, the maximum output electric power produced by this sample is about $24$ fW for $R_L=1$ M$\Omega$.

The present discussion demonstrated the high reproducibility of the BTJE, since the additional device shows results very similar to the device discussed in the main text. This shows that the bipolar thermoelectricity induced by the spontaneous PH symmetry breaking is a large and robust effect. Indeed, the presence of the effect does not require a fine tuning of the parameters when the Josephson coupling is sufficiently suppressed.

\end{document}